# PARTICLE PRODUCTION IN RELATIVISTIC NUCLEAR COLLISIONS

J. Vrláková, janka.vrlakova@upjs.sk and S. Vokál, stanislav.vokal@upjs.sk, P. J. Šafárik University, Košice, Slovakia

## INTRODUCTION

The experimental data of relativistic heavy ion interactions in photoemulsion have been studied. The fluctuations on particle production in relativistic collisions have been analysed by the scaled factorial moment method. These fluctuations may be used to signal the formation of a quark gluon plasma in the early stage of heavy ion interactions at high energies [1]. The experimental data of $^{208}Pb$ nuclei at 158 A GeV/c and $^{32}S$ at 200 A GeV/c taken with emulsion track detector have been analysed. The results have been compared with values obtained from the other experimental and model data.

## EXPERIMENT AND METHOD OF ANALYSIS

The emulsion track detector was irradiated horizontally by 158 A GeV/c $^{208}Pb$ beam at CERN. Details of this experiment can be found in [2]. Interactions of $^{32}S$ at 200 A GeV/c in emulsion have been studied also, experimental details have been published in [3]. All charged particles of measured interactions were classified according to the commonly accepted emulsion experiment terminology into groups. The group of relativistic (shower) particles includes mainly particles produced in the interactions. The relativistic particles were those with $\beta > 0.7$ emitted outside of fragmentation cone. The polar ($\Theta$) and azimuthal ($\Psi$) emission angles of all tracks have been measured. For each relativistic particle the value of pseudorapidity has been calculated as

$$\eta = -\ln\left[\tan\frac{\Theta}{2}\right]. \qquad (1)$$

A. Bialas and R. Peschanski suggested to study the dependence of factorial moment $F_q$, where $q$ is the order of the moment, as a function of the bin width $\delta\eta$ [4, 5]. The intermittent behaviour should lead to a power law dependence $F_q \propto (\Delta\eta/\delta\eta)^{\varphi_q}$, $\varphi_q > 0$, $\Delta\eta$ is the pseudorapidity interval of produced relativistic particles. The horizontal factorial moment method has been used for our analysis. Detailed description of this method can be found in [6].

## RESULTS OF ANALYSIS

Central events of $^{208}Pb+Em$ with number of relativistic particles $n_s > 350$ have been selected for analysis. The $ln <F_q>$ dependences on $ln M$ have been calculated by horizontal factorial moment method, where $M$ is number of equal bins of size $\delta\eta$ into which the pseudorapidity interval $\Delta\eta$ has been divided. Some results of this analysis have been published in our previous papers [7 - 9].

In this work we present some new results of our next analysis. The comparison with modified Cascade Evaporation (CEM) [10] and FRITIOF [11] models has been done. The dependences of values of slopes ($ln <F_q> = \alpha_q + \varphi_q . ln M$) on the order of factorial moments $q$ have been studied for groups with different degree of centrality. In Fig.1 the dependences of values of slopes ($\varphi_q$) on $q$ for experimental and model data (events with $n_s > 350$) are shown. The dependences of slopes on $q$ for events with $n_s > 1000$ are similar. The values of slopes obtained from CEM and FRITIOF are fairly smaller than those for experimental data.

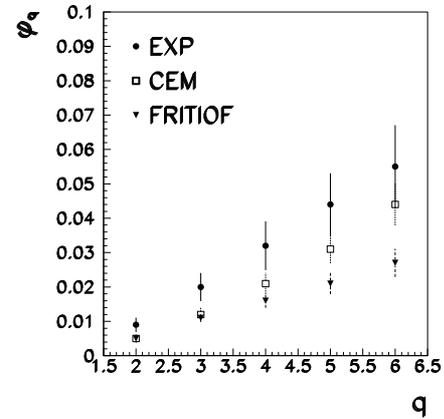

Fig.1: The $\varphi_q$ dependences on $q$ ($q=2-6$) for $Pb +Em$ (experiment, CEM and FRITIOF) with $n_s> 350$, $\Delta\eta=0-7.4$.

The interactions of $^{32}S$ at 200 A GeV/c obtained by the same standard emulsion method have been studied also. The dependences of values of slopes ($\varphi_q$) on $q$ for $^{208}Pb$ and $^{32}S$ central collisions are presented in Fig.2.

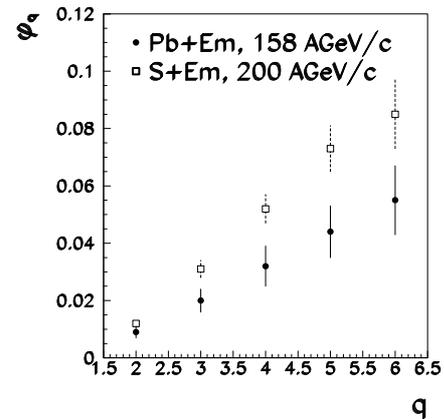

Fig.2 : The $\varphi_q$ dependences on $q$ ($q=2-6$) for $^{208}Pb$ induced interactions at 158 and $^{32}S$ at 200 A GeV/c.

The comparison with other interactions induced by the same nuclei at different momenta or nuclei with very similar mass ($^{28}Si$) has been done. The characteristics of experimental data samples – beam nucleus, momentum

($p$), total number of events ($N_T$), number of selected central events ($N_C$), pseudorapidity interval of produced relativistic particles ($\Delta\eta$) are given in Table 1. The collisions with $n_h \geq 8$ have been selected for analysis (at 4.5-14.6 A GeV/c), where $n_h$ is the number of target fragments. In order to study the central interactions with Ag(Br) nuclei, the events with $n_f > 0$ have been excluded, $n_f$ is number of projectile spectator fragments with $Z \geq 2$. Interactions with $n_s > 80$ have been selected for analysis in case of $^{32}S$ collisions at 200 A GeV/c.

TAB.1. Experimental data

| Beam nucleus | $^{28}Si$ | $^{32}S$ | $^{28}Si$ | $^{32}S$ |
|---|---|---|---|---|
| $p$[A GeV/c] | 4.5 | 4.5 | 14.6 | 200 |
| $N_T$ | 1322 | 1318 | 1093 | 1121 |
| $N_C$ | 175 | 139 | 168 | 307 |
| $\Delta\eta$ | 0-4 | 0-4 | 0-5 | 0-7.6 |

In Fig.3 the dependences of values of slopes ($\varphi_q$) on $q$ for $^{28}Si$ and $^{32}S$ at different momenta are shown. The slope values $\varphi_q$ for data at momenta of 4.5 A GeV/c are higher than values for 14.6 and 200 A GeV/c, respectively. Similar results have been obtained for $^{16}O+Em$ interactions at 4.5- 200 A GeV/c [12].

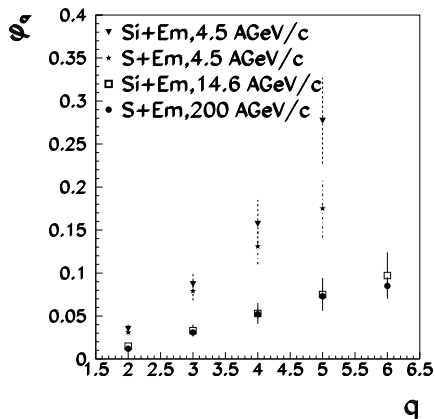

Fig.3 : The $\varphi_q$ dependences on $q$, $q=2-5$ for momenta of 4.5 A GeV/c and $q=2-6$ for momenta of 14.6 and 200 A GeV/c.

The dependence of values of slopes $\varphi_2$ on particle density per pseudorapidity unit ($\rho$) has been studied also.

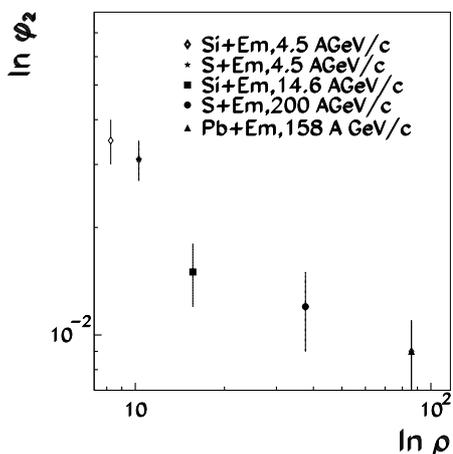

Fig.4 : The dependence of $\ln \varphi_2$ on $\ln \rho$ for different primary nuclei at 4.5-200 A GeV/c.

Our preliminary result (in logarithmic scale) obtained for bin width $\delta\eta \sim 0.1-1$ is presented in Fig.4. Similar results for other primary nuclei have been obtained in [9]. Our present results are consistent with previous chamber emulsion data [3] for pseudorapidity bin width $\delta\eta \sim 0.09 - 2$.

## CONCLUSIONS

Central interactions of $^{208}Pb$ nuclei at 158 A GeV/c and $^{32}S$ at 200 A GeV/c taken with emulsion track detector have been analysed using the horizontal factorial moment method. An evidence for the presence of intermittent behaviour has been shown. The values of slopes for experimental data are higher than values from CEM and FRITIOF models. The comparison for other primary nuclei ($^{28}Si$) at momenta of 4.5 – 200 A GeV/c has been done.

ACKNOWLEDGMENT: Financial support from the Scientific Agency of the Ministry of Education of Slovak Republic and the Slovak Academy of Sciences (Grant No. 1/0080/08) is cordially acknowledged.